\documentstyle[11pt,aaspp4]{article} 
\received{} 
\accepted{} 
\journalid{}{} 
\articleid{}{} 
\begin{document} 
\title{Plasma and Warm Dust in the Collisional Ring Galaxy VII\,Zw\,466
from VLA and ISO Observations} 
 
\author{P. N. Appleton\altaffilmark{1}} 
\authoremail{pnapplet@iastate.edu} 
\affil{Erwin W. Fick Observatory and Department of Physics and 
Astronomy, Iowa State University, Ames, IA 50011} 
 
\author{V. Charmandaris\altaffilmark{2}} 
\affil{Observatoire de Paris, DEMIRM, 61 Av. de l'Observatoire, F-75014, Paris, France } 
 
\author{C. Horellou} 
\affil{Onsala Space Observatory, Chalmers University of Technology, 
S-43992 Onsala, Sweden} 

\author{I. F. Mirabel\altaffilmark{3}} 
\affil{Service d'Astrophysique, CEA-Saclay, F-91191, Gif sur Yvette 
Cedex, France} 
\author{F. Ghigo\altaffilmark{1}} 
\affil {NRAO, P. O. Box 2, Green Bank, West Virginia WV 24944} 
 
\author{J. L. Higdon} 
\affil {Kapteyn Institute, University of Groningen, Postbus 800, 9700 
AV, Groningen, The Netherlands} 
 
\author{S. Lord} 
\affil {IPAC, MS100-22, California Institute of Technology, Pasadena, CA 91125} 
\altaffiltext{1}{Visiting astronomer at NRAO. The National Radio 
Astronomy Observatory is a facility of the National Science Foundation 
operated under cooperative agreement by Associated Universities, Inc.} 
 
\altaffiltext{2}{Marie Curie Fellow.} 
\altaffiltext{3}{also at Instituto de Astronom\'\i a y F\'\i sica del
Espacio. cc 67, suc 28. 1428, Buenos Aires, Argentina} 
Send off-print requests to P. N. Appleton e-mail: pnapplet@iastate.edu 
 
\newpage 

\begin{abstract} 

We present the first mid-infrared (Mid-IR) ($\lambda$5-15$\mu$m) and
radio continuum ($\lambda\lambda$20,~6 and 3.6\,cm) observations of the
star-forming collisional ring galaxy VII\,Zw\,466 and its host group
made with the Infrared Space Observatory\footnote{Based on
observations with ISO, an ESA project with instruments funded by ESA
Member States (especially the PI countries: France, Germany, the
Netherlands and the United Kingdom) with the participation of ISAS and
NASA.} and the NRAO Very Large Array.  A search was also made for CO
line emission in two of the galaxies with the Onsala 20m radio
telescope and upper limits were placed on the mass of molecular gas in
those galaxies.  The ring galaxy is believed to owe its morphology to
a slightly off--center collision between an ``intruder'' galaxy and a
disk. An off--center collision is predicted to generate a radially
expanding density wave in the disk which should show large azimuthal
variations in overdensity, and have observational consequences. The
radio continuum emission shows the largest asymmetry, exhibiting a
crescent-shaped distribution consistent with either the trapping of
cosmic-ray particles in the target disk, or an enhanced supernova rate
in the compressed region.  On the other hand, the ISO observations
(especially those made at $\lambda$9.6$\mu$m) show a more scattered
distribution, with emission centers associated with powerful star
formation sites distributed more uniformly around the ring. Low-signal
to noise observations at $\lambda$15.0$\mu$m show possible emission inside the
ring, with little emission directly associated with the \ion{H}{2} regions. The
observations emphasize the complex relationship between the generation
of radio emission and the development of star formation even in
relatively simple and well understood collisional scenarios.

\end{abstract} 

\section{Introduction} 

Collisions and interactions between galaxies provide a special
opportunity for astronomers to study galaxies in non-equilibrium
states (e.g. Schweizer 1997). Of particular interest are collisional
systems which have well defined initial conditions and clear
observational consequences. Models of ``head-on'' collisions, which
began with the pioneering work of Lynds \& Toomre (1976), and Toomre
(1978) and have been extended considerably in both complexity and
scope (see review by Appleton \& Struck-Marcell 1996, Struck (1997)),
fulfill those requirements. The radially expanding density waves
driven into the disk of the ``target'' galaxy by such a collision,
can not only explain the phenomenon of the collisional ring galaxy, as
Toomre had originally suggested (the ``Cartwheel'' is perhaps the most
well known example), but can also be useful as a means of exploring
other aspects of the galactic disk affected by the compression of the
ISM. In particular, in slightly off-center collisions, the expanding
wave driven by the passage of the ``intruder'' through the disk can
advance with greater strength in one direction than another, allowing
for tests of such phenomena as star formation thresholds (see Appleton
\& Struck-Marcell 1987, Charmandaris, Appleton \& Marston 1993) and
studies of the compression of plasma--such as the cosmic-ray
populations trapped in the disk.

We present radio continuum, Mid-IR and CO-line observations of a particularly well
studied northern ring galaxy VII\,Zw\,466, first discovered by Cannon
et al. (1970). As we shall show, the radio continuum emission traces
the thermal and non-thermal (relativistic) particles in the disk of
the ring galaxy, whereas the Mid-IR emission highlights unusually warm
dusty regions associated, most likely, with the neutral/molecular gas
boundaries surrounding the powerful O/B associations found in the
ring. The observations will underline the differences between the
fluid-like components of the ISM (in this case the cosmic-ray ``gas'')
and the star-formation regions that have developed as a result of the
compression by the radially expanding wave.

VII\,Zw\,466 in one of the best studied northern rings (see Thompson
\& Theys 1978) and unlike its relative in the South, the ``Cartwheel''
(see Higdon 1993, 1996; Charmandaris et al. 1999a), it comprises of a
single blue ring of star formation with no inner ring or obvious
nucleus. VII\,Zw\,466 was observed along with a small sample of other
galaxies, as part of a major study of the star formation properties
and metal content of northern ring galaxies (Marston \& Appleton 1995,
Appleton \& Marston 1997, Bransford et al. 1998-hereafter
BAMC). VII\,Zw\,466 was also mapped with the VLA in the neutral
hydrogen line and these observations provided information about the
kinematics and dynamics of the cool gas in the ring galaxy and its
nearby companions (Appleton, Charmandaris \& Struck 1996, hereafter
ACS). The \ion{H}{1} observations suggested that the ring was formed
by a collision between the progenitor of VII\,Zw\,466 and a gas-rich
dwarf spiral. A numerical model of the interaction was able to
reproduce well the \ion{H}{1} filaments extending from VII\,Zw\,466,
and a similar ``plume'' extending back from G2 towards the ring
galaxy.

In this paper we present ISO Mid-IR, VLA radio continuum, and Onsala
Space Telescope millimeter observations of VII\,Zw\,466 and its inner
group. In $S$2 we describe the details of the observations. In $S$3 we
present the intensity maps, and in $S$4 the broad-band spectral energy
distribution and radio spectral index maps of VII\,Zw\,466 and
companions.  In $S$5 we discuss the implications of the radio results,
especially in the context of the compression of the ISM and the
implications for the efficiency of the synchrotron emission from
relativistic electrons. In $S$6 we explore the consistency between
optically determined star formation rates and radio observations,
allowing us to estimate the number of UV photons heating the dust seen
in the ISO observations.  This well constrained UV flux is discussed
in the context of the ``warm'' dust distribution seen with ISO. We
attempt to develop a complete picture of VII\,Zw\,466 as a
collisionally compressed galaxy in $S$7.  We state our conclusions in
$S$8. In the Appendix we present upper limits to the molecular content
to VII\,Zw\,466 and one of its companions.
 
Throughout this paper, we will adopt a Hubble constant of 80
km\,s$^{-1}$ Mpc$^{-1}$. Using the \ion{H}{1} systemic velocity of
14,465 km\,s$^{-1}$ (ACS) we therefore assume a distance to the
VII\,Zw\,466 group of 180 Mpc.
 
\section{The Observations} 
 
\subsection{VLA Radio Continuum Observations}

The $\lambda$6\,cm (4.86 GHz = C--band) radio observations were made
with the D-array of the VLA on August 25 1992. Further observations
were made at $\lambda$20\,cm (1.425 GHz = L-band) and $\lambda$3.6\,cm
(8.43 GHz = X-band) with the C-array on October 28 1994. AIPS was used
to do the standard calibration and data reduction. Images were made
and CLEANed using the AIPS routine MX. The resulting synthesized beam
sizes for the $\lambda$$\lambda$20\,cm, 6\,cm and 3.6\,cm observations
were 18$\times$12 arcsecs, 18$\times$10 arcsecs, and 13.3$\times$8.7
arcsecs respectively. Thus, through the choice of appropriate VLA
configurations, the resolution of the $\lambda$6\,cm and $\lambda$20\,cm
observations were closely matched. The $\lambda$3.6\,cm observations were
made using natural weighting of the interferometer observations in
order to provide the maximum sensitivity to faint features.

\subsection{OSO Millimeter-Wave Observations}

In order to attempt to detect molecular gas in VII\,Zw\,466 and its
spiral companion G2, we have searched for $^{12}$CO(1-0) emission
towards those galaxies using the 20m radiotelescope at Onsala Space
Observatory, Sweden.  The observations were made in April 1999.  At
110 GHz, the telescope half-power beamwidth is 34$\arcsec$, the
main-beam efficiency is $\eta_{mb}=T_A^*/T_{mb}$=0.55.  We used a SIS
receiver and a filterbank with a total bandwidth of 512 MHz and a
resolution of 1\,MHz. During the observations, the typical system
temperature was 500\,K.  The pointing was regularly checked on nearby
SiO masers.  The pointing offsets were always below 10$\arcsec$.  The
spectra were smoothed to a final velocity resolution of
21.8\,km\,s$^{-1}$.  First-order baseline were subtracted from the
spectra.  The total on-source integration time was 400 minutes for
VII\,Zw\,466 and 90 minutes for G2. Since the galaxies were not
detected we present in an Appendix the salient upper limits to the
total molecular mass and a comparison with other work.
 
\subsection{Mid-IR Observations} 
 
VIII\,Zw\,466 was observed by ISOCAM (Cesarsky et al. 1996) on June 28
1996 (ISO revolution 224) as part of a ISO-GO observation of bright
northern ring galaxies. The galaxy was imaged through 4 filters, LW1
(4.5~ [4.0-5.0]$\mu$m), LW7 (9.62 [8.5-10.7]$\mu$m), LW8(11.4
[10.7-12.0]$\mu$m), and LW9(15.0 [14.0-16.0]$\mu$m) with on-source
time of $\sim$9.3 minutes per filter.
 
A lens, resulting in a 3$\arcsec$ pixel field of view, was used to
create a 3$\times$2 raster map in a ``microscan'' mode in one
direction.  To improve the signal to noise ratio, the ISOCAM images
were smoothed to yield a point-spread-function with a FWHM of 5.4
arcsecs.  The overall field of view was 102$\times$102 arcsec$^2$.
 
The standard data reduction procedures described in the
ISOCAM\footnote{The ISOCAM data presented in this paper was analyzed
using ``CIA'', a joint development by the ESA Astrophysics Division
and the ISOCAM Consortium led by the ISOCAM PI, C. Cesarsky, Direction
des Sciences de la Mati\`ere, CEA, France} manual were followed
(Delaney 1998). Dark subtraction was performed using a model of the
secular evolution of the ISOCAM dark currents (Biviano et
al. 1997). Because of the well-known transient response to cosmic-ray
events in the data, cosmic ray removal was performed using a
combination of multi-resolution median filtering, and time-series
glitch removal. Memory effects in the detector were corrected using
the so-called IAS transient correction algorithm (Abergel et
al. 1996). These methods and their consequences are discussed in
detail in Starck et al. (1999). The memory effects were worst in the
LW9 filter (long-wavelength filter) which was observed first (sequence
LW9, LW8, LW7 and LW1) since we had no knowledge of the source
structure prior to our target acquisition. Hence, our signal to noise
ratio is worse in this filter, reflecting the assumed arbitrary model
point source for the state of the detectors prior to our observations.
The rms noise in our ISOCAM images varied slightly over each image and
it was typically 5 $\mu$Jy/pixel in the LW1 image, 2$\mu$Jy/pixel in
the LW7, 4$\mu$Jy/pixel in the LW8 and 6$\mu$Jy/pixel in the LW9.
 
\section{The ISOCAM and VLA Imaging of VII\,Zw\,466 and its Inner Galaxy Group} 
 
In Figure 1a,d we show the B-band greyscale image from Appleton \&
Marston (1997) of the inner VII\,Zw\,466 group. On the figure we
indicate the ring galaxy, companion galaxies VII\,Zw\,466 (Cl):G1
(elliptical, hereafter G1) and VII\,Zw\,466 (Cl):G2 (edge-on disk and
likely ``intruder'', hereafter G2) and the background galaxy
VII\,Zw\,466 (Cl):B1, hereafter B1\footnote{The names of the galaxies
presented here are consistent with the names given to them by
Appleton, Charmandaris and Struck (1996), and Appleton \& Marston
(1997), and their coordinates are given in Table 1 of the current
paper.}. Superimposed over the optical image, in Figure 1a,c, we
present contour maps of the more interesting, longer-wavelength ISOCAM
images, $\lambda$9.6$\mu$m, $\lambda$11.4$\mu$m and
$\lambda$15.0$\mu$m. In Figure 1d we show contours of the $\lambda$3.6
cm radio emission, again superimposed over the B-band images of the
group.  Three of the galaxies, VII\,Zw\,466, G2, and B1 are detected
at these the longer wavelength bands. Only the elliptical galaxy G1 is
detected at $\lambda$4.5$\mu$m as an unresolved point
source (this map is not shown). Upper limits to non-detections and
values for the fluxes of the galaxies in each band are given in
Table~1.

\begin{figure} 
\figurenum{1} 
\plotone{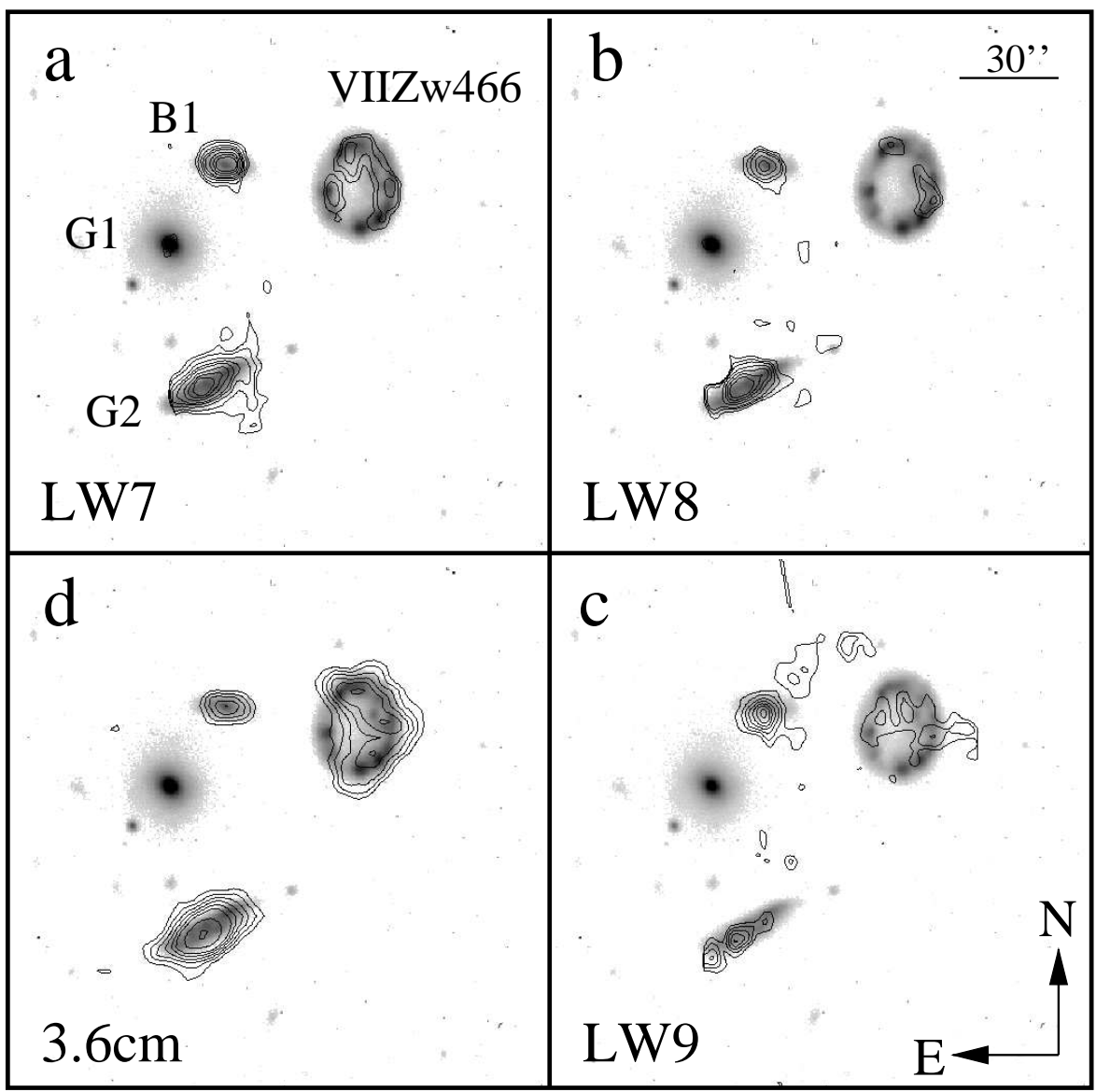}
\caption{Contours of Mid-IR emission and Radio emission
superimposed on a B-band optical image of the VII\,Zw\,466 group (a)
the $\lambda$9.62$\mu$m ISOCAM (filter LW7-bandwidth 2.2$\mu$m) image,
[contour units: 12, 15, 20, 25, 30, 40, 50 and 60 $\mu$Jy/pixel], (b)
the $\lambda$11.4$\mu$m ISOCAM (filter LW8-bandwidth 1.3$\mu$m) image
[units: 20, 25, 30, 40, 50, 60, 70 $\mu$Jy/pixel], (c) the
$\lambda$15.0$\mu$m ISOCAM (filter LW9-bandwidth 2.0$\mu$m) image
[units: 15, 20, 25, 30, 35, 40, 45 $\mu$Jy/pixels] (note that the rms
noise in this image was the highest $\sim 6\mu$Jy/pixel), (d) the
3.6\,cm radio continuum [units 60, 80, 100, 120, 140, 160, 200, 240
$\mu$Jy/beam area].\label{fig1}}
\end{figure}

\placefigure{fig1}

The distribution of Mid-IR emission at $\lambda$9.6$\mu$m and
$\lambda$11.4$\mu$m shown in Figure 1a and b are very similar except
that the signal to noise level is higher in the $\lambda$9.6$\mu$m image. At
these wavelengths, the emission in VII\,Zw\,466 is concentrated in three
main regions, one in the north, one in the east and one in the western
quadrants of the ring (the eastern source is missing in $\lambda$11.4$\mu$m
image, but this is most likely because the source is weaker and just
drops below the detection threshold at this wavelength).  The
distribution of Mid-IR emission at these wavelengths follows quite
closely the brightest H$\alpha$ sources in the ring. Figure 2a shows
the distribution of H$\alpha$ emission contours superimposed on a
grey-scale representation of the $\lambda$9.6$\mu$m emission.  Strong
emission in both bands is seen from the background galaxy B1, and the
edge-on disk G2. The emission from G2 is elongated along the major
axis of the galaxy and unresolved along the minor axis. The elliptical
galaxy (G1) is detected at $\lambda$4.5$\mu$m, and is marginally
detected at $\lambda$9.6$\mu$m. The background galaxy B1 is known to
be at approximately twice the redshift of the VII\,Zw\,466 group, and
its Mid-IR emission provides a working point-spread-function for our
observations because its emission is unresolved at all wavelengths at
which it is detected ($\lambda$$\lambda$9.6, 11.4 and 15 $\mu$m).

\placetable{table1}

In Figure 1c we show the $\lambda$15.0$\mu$m emission from the
group. This wavelength was the first observed in the observing
sequence, and is therefore of lower sensitivity than the other
observations (see $S$2). The emission from VII\,Zw\,466 is quite faint
and is only detected at the 5$\sigma$ level when averaged over the
whole galaxy. Nevertheless, the brightest emission
knot is seen centered on the western quadrant of the ring, and faint
emission is also seen from the eastern knot.  Very faint emission may
be detected from the interior of the ring in the north, but the
emission is weak and would require higher sensitivity observations to
confirm its reality.  Oddly, the northern ring-source, seen at both
$\lambda$9.6$\mu$m and $\lambda$11.4$\mu$m, is absent in these
observations. Considering that the northern knot is associated with
the brightest H$\alpha$ emitting regions, the failure to detect this
knot at $\lambda$15\,$\mu$m is surprising (see Fig. 2a), but may also reflect
the poorer quality of the $\lambda$15\,$\mu$m observation (see
discussion on memory effects in $S$2.2).

We note that the emission from the edge-on disk galaxy (G2) at
$\lambda$15$\mu$m breaks up into three discrete sources. In this
region, a line of bad detectors crossed the galaxy position just
south-east of the nucleus of G2. Flux from this region could not be
recovered by the micro-scan technique because the galaxy was too close
to the edge of the detector array. Hence the total flux for G2 in this
band is probably underestimated in our observations by up to 20\%.

\begin{figure}[ht]
\plotone{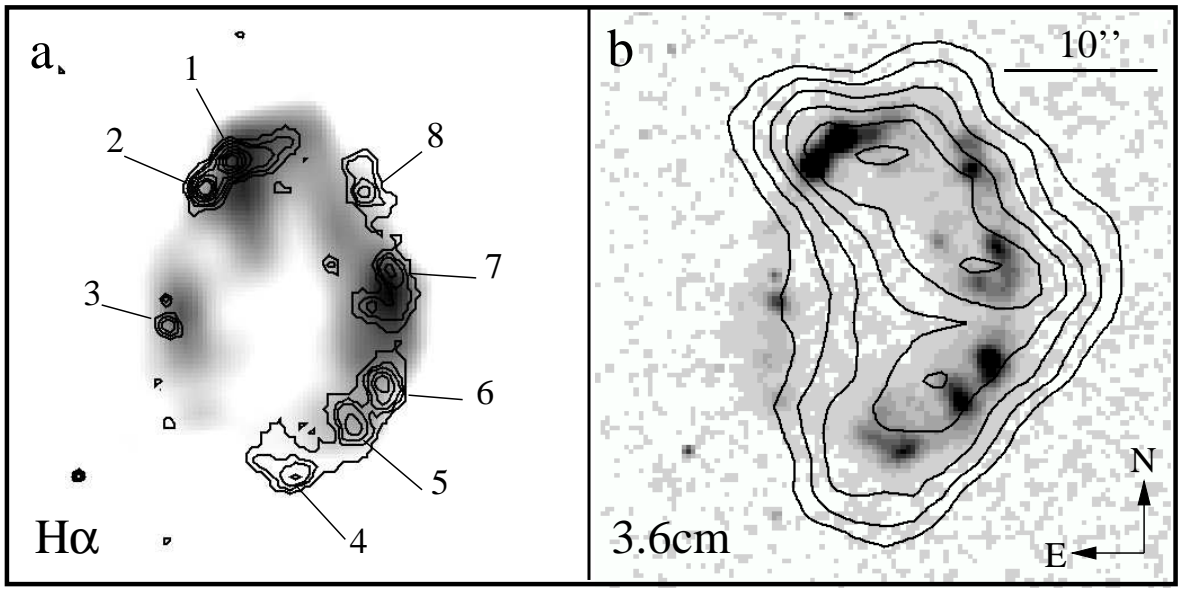}
\figurenum{2}
\caption{(a) Contours of H$\alpha$ emission superimposed on a
greyscale image of the $\lambda$9.6$\mu$m ISO image. The fluxes of the
H$\alpha$ knots numbered here are presented in Table 4 along with the
$\lambda$9.6$\mu$m fluxes. (b) Contours of 3.6\,cm wavelength radio
emission superimposed on a greyscale representation the H$\alpha$
distribution. \label{fig2}}
\end{figure}

\placefigure{fig2}

In Figure 1d, we present the radio continuum emission at $\lambda$3.6
cm. The radio emission is superficially similar to the emission at
$\lambda$9.6$\mu$m. However, it is significantly more crescent-like,
and, unlike the $\lambda$9.6$\mu$m image, shows no emission associated
with the eastern star-formation complex. The brightest radio emission
comes from the north-west quadrant and, unlike the Mid-IR maps, does
not show a close correspondence with individual \ion{H}{2} regions,
although it follows roughly the shape of the ring. There is a tendency
for the radio emission to lie on the {\it inside edge} of the ring
defined by the stars and \ion{H}{2} regions. Figure 2b shows the
$\lambda$3cm radio emission contours superimposed on a greyscale
representation of the H$\alpha$ emission. The ridge of radio emission
along the south-western edge of the ring and along the northern
regions of the ring peaks inside the ring. Only in the western ring is
there a close correspondence between the peak of the radio emission
and the peak in the H$\alpha$. We will argue from spectral index
considerations, that this is where the radio emission is predominantly
thermal.

In Figure 3a,b, we present the radio continuum maps of the
longer-wavelength 20 and 6\,cm VLA observations. Although the
observations are of lower resolution, they show the same
crescent-shaped distribution seen in Figure 1d, with no evidence for
emission from the eastern quadrant of the ring.  Radio fluxes and
globally-averaged radio spectral indices are given for VII\,Zw\,466 and
the other galaxies detected in Table~2.  The galaxy G2, the edge-on
spiral believed to be the ``intruder'' galaxy, is strongly detected
(see Figure 1d and 3a,b), and fainter emission is seen from the known
background galaxy B1. Galaxy G2 shows a flattening of its radio
spectrum as one goes to higher frequencies (Table~2), whereas the
globally average values for the spectral index are similar for the
ring galaxy and the background galaxy B1. The steeper values of the
spectral index are typical of a predominantly synchrotron spectrum
whereas, for G2, the flattening of the spectrum may be a result of an
increasingly thermal component, perhaps resulting from a powerful
starburst.
 
\begin{figure}[ht]
\plotone{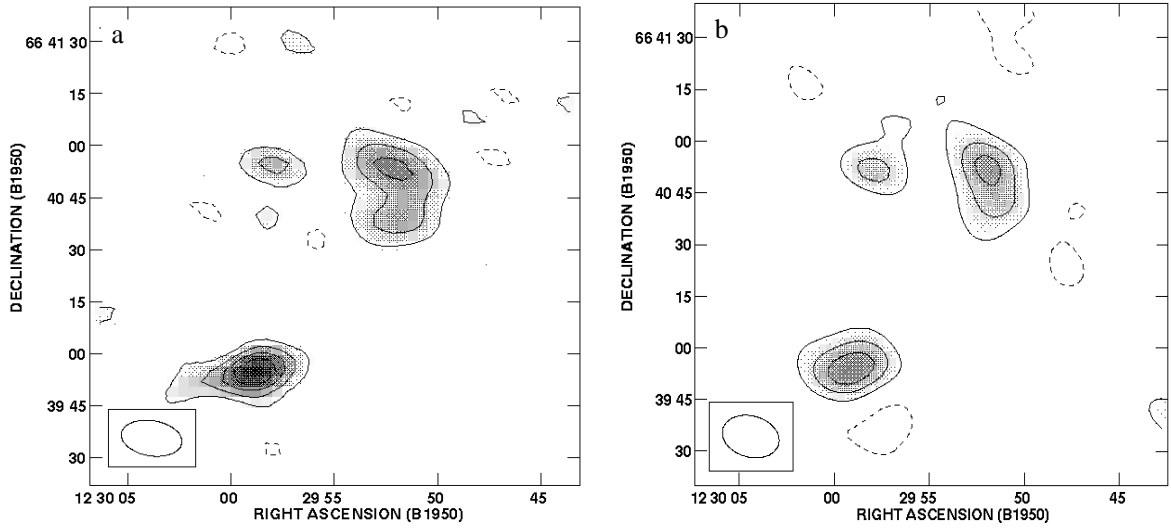} 
\figurenum{3} 
\caption{(a) 6\,cm radio map of the inner VI~Zw~466 group.  Contour
levels are -1, 1, 2, 3 $\times$ 85 $\mu$Jy/beam, (b) 20\,cm map of the
VII\,Zw\,466 group. Contour levels are -1, 1, 2, 3 $\times$ 290
$\mu$Jy/beam. \label{fig3}}
\end{figure}

\placefigure{fig3}
\placetable{table2}

In Figure 4a and b, we present the radio spectral index maps of the
galaxies (computed between $\lambda$$\lambda$20 and 6\,cm and between
$\lambda$$\lambda$6 and 3.6\,cm). Although the resolution in these
maps is poor (we convolved the higher frequency observation to the
resolution of the lower frequency one in each case), the ring galaxy
shows some structure in these maps. The ``horn'' of the crescent in
both maps show radio spectral indices typical of synchrotron emission,
with a value of $\alpha$ = -0.6 to -0.7 in both maps (a value close to
the average for the whole galaxy-Table~2).  However, in the center of
the ``crescent'', which would correspond to the western ``hotspot'' in
Figure 1d in the ring, the radio spectrum shows a change of slope from
a spectral index which is quite steep between $\lambda$20 and 6\,cm,
$\alpha$ = -1.0, to a flat spectrum at the shorter wavelengths of
$\lambda$6 to 3cm ($\alpha$ = 0.0). Such a change in slope is typical
of a radio spectrum which shifts from one dominated by synchrotron
emission to a thermal spectrum. The radio observations suggest that
there is a strong source of hot plasma in the western part of the
ring. Optically, this region does not show strong H$\alpha$ emission,
but lies between two moderately bright \ion{H}{2} regions.

\begin{figure}
\plotone{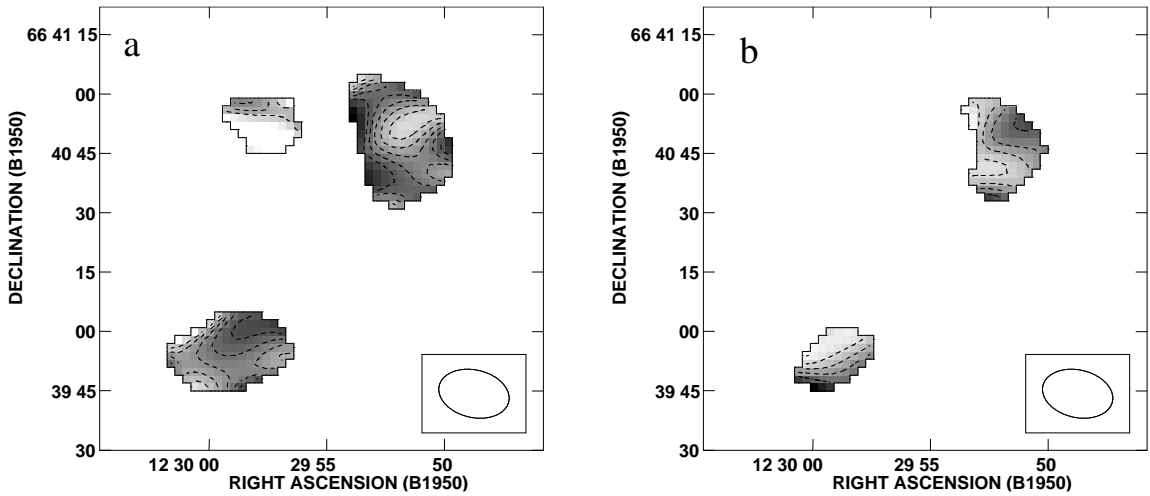}
\figurenum{4}
\caption{The radio spectral-index maps for the VII\,Zw\,466
group (using the convention S $\sim$ $\nu$$^{\alpha}$) (a)
6\,cm/20\,cm : range of spectral index is from white (-1.0) to black
(-0.6), the contour interval is 0.1. Note that at these longer
wavelengths the mid-point of the radio ``crescent'' has a steeper
spectrum than the ``horns'', (b) spectral index map based on the
ratios of the 3\,cm/6\,cm data; white corresponds to -0.6 and black
corresponds to a spectral index of 0, with a contour interval of
0.2. The flat spectrum at the high frequencies is typical of a thermal
emission. The spectral index in the ``horns'' of the crescent is about
the same in both figures, around -0.6 to -0.7.\label{fig4}}
\end{figure}

\placefigure{fig4}

\section{Spectral Energy Distributions and the origin of the Mid-IR 
emission} 

In Figure 5 we present spectral energy distributions (SEDs) for the
four galaxies observed, based on our previous optical/IR photometry
(Appleton \& Marston 1997), the low-resolution Faint Source Catalog of
IRAS (Moshir, Kopman \& Conrow 1992), and the current radio data. The
detection of emission in the Mid-IR is an order of magnitude lower
than the previous upper limits to the $\lambda$12$\mu$m flux from
IRAS.  The emission from the elliptical companion G1 at
$\lambda$4.5$\mu$m and $\lambda$9.6$\mu$m is quite consistent with the
emission being an extrapolation of the stellar continuum seen at
shorter wavelengths. It is known that the Mid-IR spectrum of
elliptical galaxies is dominated by their evolved stellar population,
and can be modeled fairly well using a blackbody continuum with a
temperature of$\sim$\,4500K (Madden 1997). This is not the
case for the other galaxies where there is a clear Mid-IR excess.

\begin{figure}
\plotone{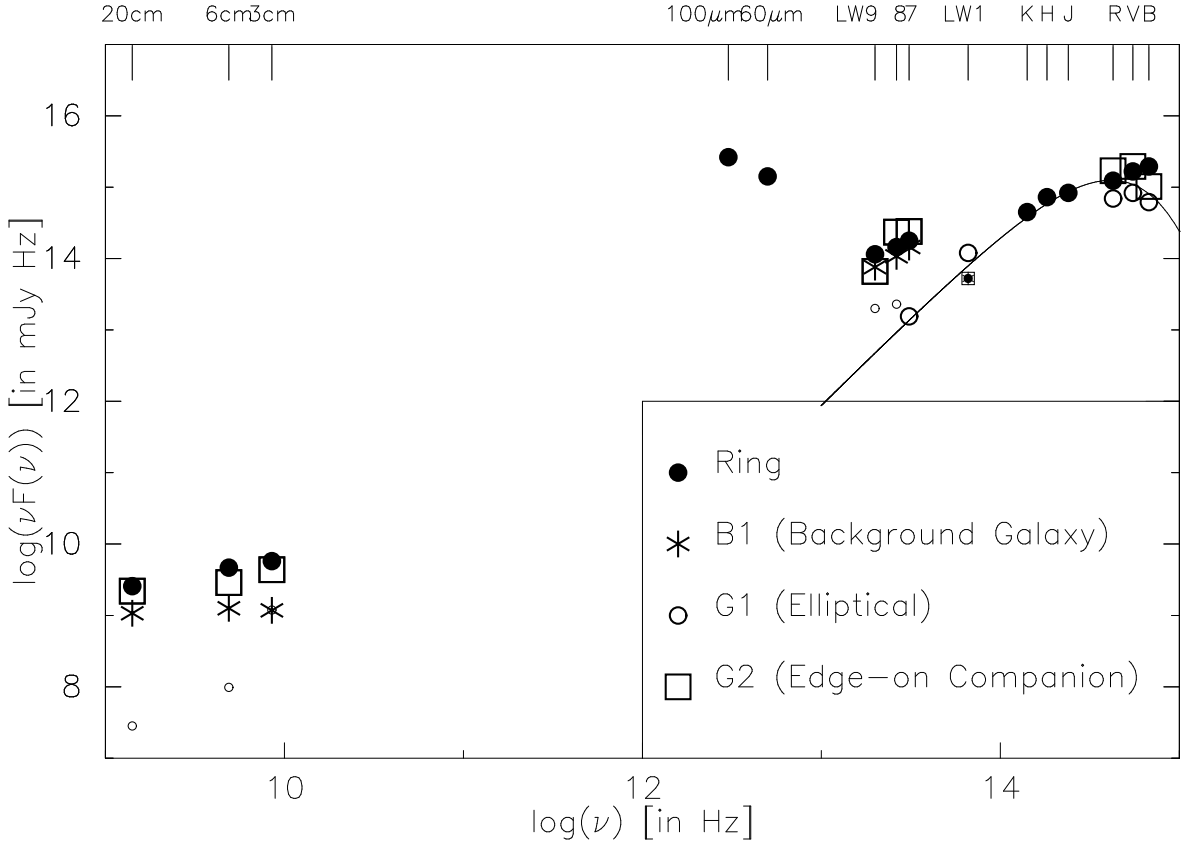} 
\figurenum{5} 
\caption{Spectral Energy Distribution of the four galaxies from the
optical to radio. The very small open circles refer to upper limits to
the elliptical G1. We show for reference, a black-body spectrum
corresponding to a temperature of 4000\,K, passing through the R-band
point for normalization. In all but the elliptical galaxy, G1, the
Mid-IR shows an excess not expected from an extrapolation of the hot
stellar continuum from the galaxies.\label{fig5}} 
\end{figure}

\placefigure{fig5}

Although we explored a pure dust model of the SED of the Mid-IR
emission from VII\,Zw\,466 and obtained a satisfactory fit to these
data with a dust temperature of $T_D$=226 K, we have decided not to
present these results here. It is very likely that the spectrum is
affected by emission from Unidentified Infrared Bands (UIBs).  These
broad lines (centered at $\lambda$$\lambda$ 6.2, 7.7, 8.6 and
11.3$\mu$m) are commonly seen in active star formating regions, and
are usually attributed to emission from PAHs, since they can be
produced by the stretching of C-C and C-H bonds (L\'eger \& Puget
1984; Verstraete, Puget \& Falgarone, 1996). As a result, we find that our LW7
(9.6$\mu$m) and LW8(11.4$\mu$m) filters are likely to be significantly
affected by the 7.7 and 11.3$\mu$m UIBs, after taking into account the
redshift of the galaxies. Indeed it is likely that in these bands,
much of the emission we see originates in the UIB features.

In order to facilitate a comparison between our observations and other
star formation active regions in which UIBs are present, we show in
Figure 6, LW7, LW8 and LW9 fluxes for VII\,Zw\,466 superimposed on two
contrasting spectra taken with ISO of the well known Antenna galaxy
(Mirabel et al. 1998, Vigroux 1997). The spectra have been redshifted
to that of VII\,Zw\,466.  It can be seen from Figure 6 that the UIBs
at $\lambda$7.7 and $\lambda$8.6 $\mu$m are partially shifted into the
LW7 ($\lambda$9.6$\mu$m) filter. Similarly, the rather brighter UIB at
$\lambda$11.3$\mu$m, if present, is expected to affect the LW8
($\lambda$11.4$\mu$m) flux measurement. Because of the redshift of
VII\,Zw\,466, the LW9 bandpass (centered on $\lambda$15.0$\mu$m) falls
between the [NeII] and [NeIII] emission lines and effectively gives an
uncontaminated measurement of the thermal continuum at
$\lambda$15$\mu$m.  It can be seen from a comparison between our flux
measurements for VII\,Zw\,466, and the Antenna spectra, that our
measurements are consistent with a spectral energy distribution
similar to Knot B in the Antenna. The main difference between Antenna
Knots A and B is the strong enhancement in the $\lambda$12-16$\mu$m
thermal continuum, which is very powerful in Knot A as compared with
the strength of the UIBs. In Knot B, the contributions to the spectrum
from the UIBs and a mild thermal continuum are much more comparable.

\begin{figure}
\plotone{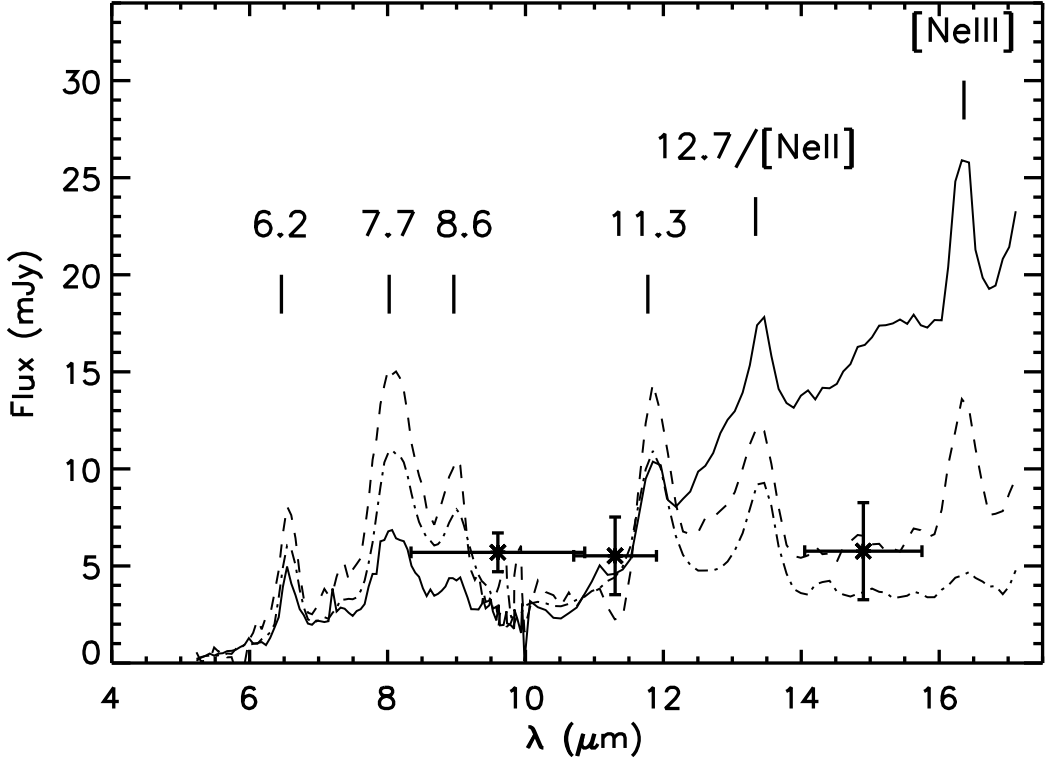}
\figurenum{6}
\caption{The flux measurements for VII\,Zw\,466 (crossed plus error bars)
compared with the Mid-IR spectra of three regions in the Antennae
galaxy (Knot A--solid line, Knot B--dotted line and Knot Z dot--dashed
line) which have been normalized to the $\lambda$11.4$\mu$m (LW8) flux
of VII\,Zw\,466. The vertical bars shows the 3$\sigma$ error on the
flux measurements, dominated by the uncertainty in the Mid-IR
background in the observations. The horizontal ``error-bar'' shows the
width of the filters used. The location of the UIB Mid-IR features and
the [NeII] and [NeIII] emission line features corrected for the
redshift of the galaxy are also indicated. Note that the $\lambda$9.6
and 11.4$\mu$m filters would be expected to be contaminated by UIB
features, if present, but the $\lambda$15$\mu$m filter would fall
bellow the [NeIII] emission lines. \label{fig6}}
\end{figure}

\placefigure{fig6}

In order to further quantify this similarity, we present in Table~3
the LW8/LW7 and LW9/LW7 flux ratios that would be expected for various
input spectra obtained from publicly available ISO data sets. We take
four different input spectra, three from the Antenna galaxy (Knots A,
B and Z from Mirabel et al. 1998 and Vigroux et al. 1997), and one
from the ISO Mid-IR spectrum of Arp 220, which shows a very strong
silicate absorption feature (Charmandaris et al. 1999b).  Table~3
shows that the flux ratios most consistent with the observed ratios
for VII\,Zw\,466 is Knot B of the Antenna.  This is a result of the
contaminating influence of the $\lambda$11.3 $\mu$m UIB relative to
the $\lambda$7.7 and $\lambda$8.6$\mu$m UIB features. In most other
cases the ratio of the LW8/LW7 flux is significantly greater than
unity due to the combination of both the strength of the
$\lambda$11.3$\mu$m UIB feature and a relatively strong rising thermal
continuum beyond $\lambda$10$\mu$m. An extreme case is Arp 220, where the
LW9/LW7 ratio is large, reflecting a combination of a rapidly
increasing thermal continuum at longer wavelengths and a strong dip in
the spectrum around $\lambda$10$\mu$m due to silicates seen in
absorption\footnote{We also compared our results with spectra of the
rare case of sources in which silicates are seen in emission, rather
than absorption (This is the case for some unusual \ion{H}{2} regions
observed recently by ISO--J. Lequeux, Private Communication). Again,
this did not yield the ratio of LW8/LW7 emission seen in
VII\,Zw\,466.}.

\placetable{table3}

Table~3 shows that the Mid-IR spectrum of the G2 and the background
galaxy B1 are similar to VII\,Zw\,466 in their ratios of LW8/LW7 and
LW9/LW7. In the case of G2, the latter ratio is not reliable because
of the uncertain flux of G2 in the LW9 band (see $S$3).

\section{Explaining the Radio Morphology: Compression of the ISM
and/or Increased Supernova Rate?} 

\subsection{Amplification and Trapping of Cosmic-Ray Particles in a Compressed Disk?}

Models of slightly off-center collisions between galaxies lead to the
prediction that the ISM will be compressed more on one side of the
target galaxy compared with the other (Appleton \& Struck-Marcell
1987). This effect was used by Charmandaris, Appleton \& Marston
(1993) to explore threshold star formation behavior in Arp 10. In
VII\,Zw\,466, the crescent-shaped distribution of the radio emission
at $\lambda$3\,cm is very suggestive of both compression and trapping
of relativistic particles in the disk.

In order to explore this idea further, we will briefly explore the
possibility that the crescent-shaped enhancement in the radio emission
can be explained in terms of a modest (approximately 5-10 times)
compression of the galactic disk on one side of the galaxy compared
with the other. Such density enhancements are similar to those
predicted in the model by ACS used to explain the \ion{H}{1} fingers
extending from VII\,Zw\,466.

Helou \& Bicay (1993) showed that the radio synchrotron luminosity of
a relativistic electron in a galactic disk was:

\begin{equation}
L_{synch}=\chi~L_{cr}~(t_x/(1+t_x))
\end{equation}

where L$_{cr}$ is the luminosity in the cosmic ray population, and
t$_x$ is the ratio of two critical time-scales,
t$_x$=t$_{esc}$/t$_{synch}$. Here t$_{esc}$ is the time taken for a
cosmic ray electron to diffuse vertically within the disk to a point
where it can escape confinement, t$_{synch}$ is the synchrotron
lifetime of an electron in an average galactic magnetic field, and
$\chi$ is a constant close to unity. To extract the maximum available
luminosity from the electron before it looses too much energy,
t$_{esc}$ $>>$ t$_{synch}$ (or t$_x$ $>>$ 1) and then L$_{synch}$
$\sim$ L$_{cr}$.  On the other hand, if t$_{esc}$ $<$ t$_{synch}$ (or
t$_x$ $<$ 1), a result of the electron escaping from the disk before
it can deposit all its energy, then L$_{synch}$ $<$ L$_{cr}$ and the
radio synchrotron luminosity will be significantly reduced. Helou \&
Bicay estimate the values of the two time-scales as follows:

\begin{equation}
t_{esc}\simeq 10^7 [h_{disk}/1kpc]^2[1pc/l_{mfp}](cos\phi)^{-1}~~~yr
\end{equation}
and
\begin{equation}
t_{synch} \simeq 8 \times 10^9 [B/1\mu G]^{-2}[E_o/1GeV]^{-1} (sin\phi)^{-2}~~~yr
\end{equation}

Here, h$_{disk}$ is the scale-height of the magnetically confined
galactic disk, l$_{mfp}$ is the mean free path of the electrons in the
disk, B is the magnetic field strength, $\phi$ is the angle of the
electron momentum vector to the magnetic field vector, and E$_o$ is
the initial injection energy of the particle. The escape lifetime is a
random walk out of the disk by scattering off magnetic irregularities
which set the scale of l$_{mfp}$. Putting in realistic values for the
above parameters (h$_{disk}$~=~1~kpc, l$_{mfp}$~=~1~pc, B = 5~$\mu$G,
E$_o$ = 5 GeV) we find that t$_{esc}$ = 10$^7$ yrs and t$_{synch}$ = 6
x 10$^7$ yrs, and so for cosmic ray electrons with
energies of a few GeV, t$_{esc}$ is shorter than the synchrotron
lifetime, and we are in the regime in which L$_{synch}$
$<L$$_{cr}$\footnote{In our galaxy, the break in the cosmic ray
electron spectrum (see discussion by Longair 1994), in the range 1-10
GeV is most likely a result of the escape in the electrons from the
Galactic disk at the lower energies--hence the most efficient
synchrotron production occurs for those that remain trapped--with
energies $>$ 5 to 10 GeV}.

We can now ask the question--what happens if the disk containing a
steady-state cosmic-ray population of relativistic electrons before
the collision, is compressed by a factor of 5-10 in part of the
ring-wave for about 50-100 million years? This compression timescale
comes from the width of the stellar ring (3.6 arcsec or approximately
3\,kpc), divided by the radial expansion velocity of the ring
V$_{rad}$ = 32\,km\,s$^{-1}$ based on \ion{H}{1} observations (see
ACS). The compression has a two-fold effect. Firstly, it is likely
that the component of the magnetic field perpendicular to the
compressional shock wave will be amplified (B proportional
to $\rho$), resulting in a decrease in the synchrotron lifetime by 25
to 100.  Secondly, the compression will also order the random
component of the magnetic field and reduce the average distance
between magnetic irregularities in the disk in a linear way,
increasing the time needed for the electron to escape from the disk by
a factor of 5 to 10. The net effect is to increase t$_x$ by a few
hundred, perhaps enough to allow the synchrotron emission from our
canonical 5 GeV electron to reach its maximum efficiency L$_{synch}$
$\sim$ L$_{cr}$. The compression of the disk in the ring wave
temporarily decreases the energy at which disk CR-electrons will be
just trapped by the disk from typically E$_{o}$ $\sim$ 10 GeV to
around 1~GeV. The effect on the overall synchrotron luminosity will be
a significant increase in the radio luminosity in the region just
behind the peak compression of the wave. In the models of the
off-center collisions, just such a crescent-shaped compression is
expected, and the fact that the radio emission is off-set a little to
the inner edge of the ring is confirmation that the basic idea is
correct. One (rather untestable!) prediction of our observations would
be that if it was possible to actually measure the electron energy
spectrum of cosmic rays in the enhanced region of VII\,Zw\,466, a
spectral break should occur at lower energies (more like 1~GeV) than
that seen in our own Galactic disk (5-10~GeV; e.g. Longair 1994).

\subsection{Supernova Rates in VII\,Zw\,466: An Alternative to Compression?}

In the last section we explored the possibility that disk
compression alone is responsible for the crescent-shaped radio
distribution. An alternative explanation for the enhanced radio
emission on the western half of VII\,Zw\,466 might be the generation
of new CR particles from Type II SN resulting from the triggered star
formation in the ring. Although this does not contradict the analysis
in the previous section (we did not say where the CR electrons
originated), it does mean that the relativistic electrons do not need
to pre-date the collision, but can be created $\it{in~situ}$ as a
direct consequence of the massive star formation process itself. In
this picture, the compression need not be the main factor responsible
for the enhanced radio emission, but rather that enhanced star
formation would lead to a higher SN rate on one side of the galaxy
compared with the other.

Condon \& Yin (1990) provide an estimate of the supernova rate from
the non-thermal component of the radio synchrotron luminosity using
the formula:

\begin{equation}
L_{NT}(W Hz^{-1}) = 1.3\times10^{23} (\nu/1GHz)^{-{\alpha}} R_{SN}~~~yr^{-1}
\end{equation}

where $\nu$ is the radio frequency of the observation, $\alpha$ is the
spectral index, and R$_{SN}$ is the rate of Type II supernova per
year. Assuming that 70\% of the radio emission from VII\,Zw\,466 is
non-thermal at $\lambda$6\,cm (a figure based on a simple fit to the
spectrum of the galaxy), we find that L$_{NT}$= 4.4 $\times$
10$^{21}$~W Hz$^{-1}$. With this figure, we find a supernova rate
R$_{SN}$ = 0.048 yr$^{-1}$. How reasonable is this number? One way to
check it, is to see whether this supernova rate agrees with the
expected rate based on our knowledge of the star formation rates in
VII\,Zw\,466. The total H$\alpha$ flux from VII\,Zw\,466 is 7.2
$\times$ 10$^{41}$ ergs\,s$^{-1}$ or 7.2 $\times$ 10$^{34}$ W (Marston
\& Appleton 1994). Based on the models of Kennicutt (1983) we can
calculate from the H$\alpha$ luminosity the rate of star formation for
stars more massive than 10 M$\odot$ to be 0.97 M$\odot$/yr for
VII\,Zw\,466 as a whole. If we assume that all stars more massive than
10 M$\odot$ explode as supernova, then the number of stars being born
per year with masses $>$ 10 M$\odot$ will be equivalent to the Type II
supernova rate. The rate will increase with time since the main
contribution to the supernova rate comes from the more populous lower
mass stars which have longer main sequence lifetimes. However, this
rate will be reasonable if it is averaged over a few x 10$^7$ yrs,
which is also the approximate propagation time of the ring wave
through a given regions of the gas disk. For an assumed upper mass
cut-off of 80 M$\odot$ and an IMF slope of -2.35 (Salpeter) the rate
of formation of massive stars is 0.045 $\times$ dM/dt or 0.044
stars/yr. This would correspond to the supernova rate (R$_{SN}$ =
0.044) under the assumptions given above. This figure is in close
agreement with the value derived from the non-thermal radio
emission. However the agreement is fortuitous. For example, a change
in slope of the IMF to -3 (a bottom heavy IMF) would increase the
supernova rate to 0.053/yr. And so a realistic value for the Type II
supernova rate derived from the H$\alpha$ fluxes would be R$_{SN}$ =
0.045 $\pm$ 0.01, or about 1 supernova every 20 years. The lower-limit
to the star formation rate implied by the H$\alpha$ observations
(uncorrected for extinction) is therefore capable of delivering the
observed radio flux through {\it in situ} Type II supernova associated
with the star-forming regions. The agreement also suggests the
H$\alpha$ flux is relatively free of absorption in VII\,Zw\,466. These
facts make VII\,Zw\,466 an interesting place to look for supernova
explosions at optical and radio wavelengths, and we would urge that
the galaxy be monitored every few years for signs of explosions in the
outer ring\footnote{We also note that we get a very similar result for
R$_{SN}$ = 0.046 from a composite optical/radio method (see equation
12 of Condon \& Yin 1990). Since this method depends not only upon the
H$\alpha$ flux, but also on the ratio of the thermal to non-thermal
radio emission, the agreement between that result and the one
presented in the main text here is confirmation that we have correctly
separated the thermal from the non-thermal radio emission (70\% to
30\%) in VII\,Zw\,466. The value of R$_{SN}$ derived by this method is
sensitive to the assumed ratio of thermal to non-thermal emission.}.

\section{The Thermal Ultraviolet Continuum}

The Mid-IR observations suggest that dust grains are being heated by
the hot young stars. What limits can be placed on the strength of the
UV emission? We are in the fortunate position in VII\,Zw\,466, of
having two relativity independent method of measuring the UV
continuum, namely the radio and the optical hydrogen recombination
line flux (The Far-IR flux can also provide a third rough measure of
the UV flux).  The first measure is from the H$\alpha$ flux, which can
be used to estimate the strength of the Lyman continuum. From the
total H$\alpha$ flux (Marston \& Appleton 1995) we can estimate the
total number of ionizing UV photons
N$_{uv}$=L$_{H{\alpha}}$/1.36$\times$10$^{-19}$ =
5.3$\times$10$^{53}$\,s$^{-1}$ (assuming an electron temperature of
10$^4$\,K).  Since no internal reddening correction has been made to the
H$\alpha$ flux, this UV flux is a lower-limit.  Secondly, we can make
a crude decomposition of the radio flux into thermal and non-thermal,
based on the spectral index radio maps and use this to estimate the UV
emission.  Approximately 30\% of the total radio flux at
$\lambda$6\,cm is estimated to be thermal. Hence, again from Condon \&
Yin, we can determine the number of UV photons from:

\begin{equation}
N_{uv} = 6.3\times10^{32} L_T(W Hz^{-1})\nu ^{0.1}(GHz) 
T_4^{-0.45}~~s^{-1}
\end{equation}

From the 6\,cm radio flux we obtain a value for N$_{uv}$ =
8$\times$10$^{53}$ if T$_4$=1 (the electron temperature in units of
10$^4$\,K).

Hence there is excellent agreement between the H$\alpha$ determination
and the radio determination for the number of ionizing photons. The
radio result suggests an upper limit to the optical extinction near
$\lambda$6563\AA~of $<$ 0.4 mag.  This would translate to A$_v$ $<$ 1
mag for a Galactic extinction curve and is very consistent with a low
value for A$_v$ of 0.2 for all of the \ion{H}{2} regions observed
spectroscopically by BAMC, except for one knot inside the ring on the
western side)\footnote{The Far-IR flux provides an additional check on
the UV flux if we assume that the FIR luminosity contains a large
component of its luminosity from re-processed UV radiation from star
forming regions.  The total FIR luminosity for VII\,Zw\,466 can be
estimated from the IRAS flux, and L$_{FIR}$= 7.99 $\times$ 10$^{36}$
W. (VII\,Zw\,466 is F12297+6641 in the IRAS Faint Source Catalog and
its log(FIR)=-13.69 W m$^{-2}$). This compares with a UV luminosity
derived from the radio estimate above of L(radio)$_{uv}$ of
2.7$\times$10$^{36}$ W, if the average UV photon energy is 20 eV, the
temperature characteristic of a typical 50,000\,K O star.}.

\section{Building A Complete Picture of the Activity in VII\,Zw\,466}

Previous observations of VII\,Zw\,466 have shown that the outer ring
consists of star clusters and H$\alpha$ emission regions consistent
with a young (10-20 million year old) population of stars recently
laid down by the passage of a radially expanding wave through the disk
(BAMC, ACS). In this paper we have presented some new ingredients,
namely the radio continuum emission and the Mid-IR emission from
ISO. We have determined that the thermal component of the radio
emission is in good agreement with the global H$\alpha$ emission in
providing a quantitative estimate of the ultraviolet continuum from
the star formation. This observation allows us to rule out the
possibility that VII\,Zw\,466 is heavily obscured by dust at optical
wavelengths.  It is this same UV continuum which is presumably
responsible for the heating of the Mid-IR emission regions seen by
ISO, at least in the $\lambda$9.6 and $\lambda$11.4\,$\mu$m ISOCAM
bands. The UV continuum must also heat the grains responsible for the
Far-IR emission, and we have seen that the fluxes we have calculated
are at least comparable with the total Far-IR flux for reasonable mean
UV photon energies. Based on the fluxes given in Table~1, we can
estimate the total Mid-IR luminosity of VII\,Zw\,466 L$_{MIR}$ =
2.39$\times$10$^{35}$\,W, which is approximately 10\% of the estimated
available UV luminosity (L$_{uv}$=N$_{uv}$$<$E$_{uv}$$>$, where we
assume the average UV photon energy $<$E$_{uv}$$>$ = 20 eV).

The distribution of \ion{H}{2} regions in VII\,Zw\,466 have already
been shown in Figure 2a and 2b. The observations, first presented by
Marston \& Appleton (1995), are further analysed here. In Figure 2a we
label the knots and present the H$\alpha$ and $\lambda$9.6$\mu$m
fluxes in Table~4.  As mentioned earlier, the 9.6$\mu$m
emission follows closely the \ion{H}{2} region distribution. Even Knot
3, the eastern knot, is detected at $\lambda$9.6$\mu$m. From Table 4
it can be concluded that the fluxes of the ISO emission do not scale
linearly with the fluxes of the \ion{H}{2} region complexes. Since
significant optical extinction is not responsible, it is more likely
that the variations from one region to another are due to an irregular
filling factor for the dust in the UV radiation field (we know that
only 10\% is absorbed on average, but this could vary wildly from one
region to another). Alternatively, if the grains responsible for the
emission are small, they may be subject to thermal spiking by single
UV photons. In this case, there is no reason to expect that the IR
flux from a collection of such grains would scale linearly with the
incident UV flux.  Figure 2b also emphasizes the difference between
the radio and H$\alpha$ distribution, which does not follow the
\ion{H}{2} regions as faithfully as the $\lambda$9.6$\mu$m
emission. This may be because the radio emission is following the
region of maximum compression of the hydrodynamic disk (see $S$5),
whereas the star formation is more stochastic showing hot-spots as
different star clusters form at slightly different times around the
ring. The observations show that, even in a galaxy in which many of
the overall parameters of the collision are quite well known, the
distribution of the various phases of the ISM and the star formation
which develops is quite complex.

\placetable{table4}

An unexpected aspect of the ISOCAM observations is the distribution of
$\lambda$15$\mu$m emission in the ring galaxy (Figure 1c--from the
ISOCAM LW9 filter). Only the H$\alpha$ knots 3 and possibly 6 (see
Figure 2a for labeling of knots) are coincident with corresponding
knots at $\lambda$15$\mu$m. It is not clear if this is because of the
poor signal to noise of this ISO band, or that the emission really
does avoid the sources of UV radiation. This result is consistent with
the majority of the \ion{H}{2} region complexes in the Cartwheel
galaxy (Charmandaris et al. 1999) which are also absent at
$\lambda$15$\mu$m, but are detected at shorter wavelengths.  Only one
extremely powerful \ion{H}{2} region complex in the Cartwheel was
detected at $\lambda$15$\mu$m, and it was suggested that this was
because of its unusually large H$\alpha$ luminosity. It is possible
that strong winds and radiation pressure may lift the grains
responsible for the $\lambda$15$\mu$m emission to large distances from
the \ion{H}{2} regions causing them to radiate in the Far-IR, unless
the \ion{H}{2} regions are especially luminous. None of the \ion{H}{2}
regions in VII\,Zw\,466 are comparible in luminosity with the very
bright complex in the Cartwheel (see BAMC).  However, we caution that
the emission from VII\,Zw\,466 in the LW9 filter is weak, and this
filter was more prone to memory effects than the shorter wavelength
filters.  Higher S/N ratio observations will be required to confirm
the above results.

\section{Conclusions} 
 
The observations at radio and IR wavelengths have shown that: 
 
1) Emission from the ring galaxy is detected in three ISOCAM bands
($\lambda$$\lambda$9.6, 11.4 and 15 $\mu$m--filters LW7, LW8 \& LW9).
The total mid-IR luminosity L$_{MIR}$ is approximately 10\% of the
available UV luminosity from star formation found by two independent
methods. The emission at $\lambda$9.6 and $\lambda$11.4$\mu$m follows,
but does not precisely scale with, the flux of the H$\alpha$ emitting
regions. This is consistent with the dust grains being heated in
clumpy, somewhat irregular distributions surrounding the \ion{H}{2}
region complexes, or that the grains are distributed uniformly, but
are small thermally-spiked grains.  The $\lambda$15$\mu$m emission is only
marginally detected, but seems to be poorly correlated with the
\ion{H}{2} regions and some of its emission may lie inside the ring.

2) The Mid-IR emission spectral energy distribution is most consistent
with emission from grains warmed by young stars. The shorter
wavelength bands observed around $\lambda$8-11$\mu$m are contaminated by
emission from Unidentified Infrared Bands (probably thermally spiked
PAHs), whereas the longer wavelength band at 15$\mu$m shows a weak
thermal continuum similar to Knot B in the Antennae galaxies.

3) Radio observations, made at $\lambda$$\lambda$3, 6 and 20\,cm,
reveal a crescent-shaped distribution of emission which peaks on the
inside edge of the ring, but is not as closely associated with the
\ion{H}{2} region complexes as the Mid-IR emission. Spectral index
variations around the ring suggest that about 30\% of the emission at
$\lambda$6\,cm is thermal, and is consistent with the observed H$\alpha$
flux. This suggests that little optical extinction is present in the
galaxy. The non-thermal component of the emission, which dominates the
radio emission, is a synchrotron component which we associate with a
cosmic ray (CR) population trapped in the disk of VII\,Zw\,466. The
enhancement in the radio flux in one side of the ring (giving it an
apparent crescent-shaped distribution) is caused by either: i) a
compressional amplification and trapping of the synchrotron emitting
particles by the compression wave which is stronger on one side of the
galaxy than the other (a result of the off-center nature of the
collision), or ii) an enhancement in the number of Type II supernova
on one side of the galaxy from the other.  Both mechanisms seem
plausible and both may play a role in defining the radio morphology.

4) By a number of different methods we calculate the Type II supernova
rate in VII\,Zw\,466 to be R$_{SN}$ = 0.045 $\pm$ 0.1 (approximately 1 every
20 years) in the ring. A comparison between optical and radio
observations suggests that VII\,Zw\,466 has unusually low optical
extinction (a result also suggested by optical
spectroscopy-BAMC). Hence this galaxy would be an ideal candidate for
automated supernova searches and for repeated radio imaging for radio
supernova.

5) The difference in morphology between the radio and optical/Mid-IR
distribution may result from the different way in which various
components of a galaxy respond to the compressional wave which is
expected to pass through the ``target'' disk in a ring-galaxy
collision. The strength of the radio emission may respond directly to
the overdensity and compression of the ISM (especially the cosmic-ray
``fluid''), whereas the star formation and associated dust is more
stochastic, resulting from parts of the disk being pushed into a
star-formation mode in a non-uniform way, leading to a
more scattered distribution of emission centers around the ring.

6) The galaxy G2, an edge-on disk a few diameters away from the ring,
is strongly detected at Mid-IR and radio wavelengths, and has a
similar spectrum to the ring galaxy. It is likely that the star
formation activity in that galaxy has been enhanced because of the
interaction with VII\,Zw\,466. Previous \ion{H}{1} observations and
modelling have suggested that G2 is the ``bullet'' responsible for the
formation of the ring in VII\,Zw\,466.
 
\acknowledgments

P. N. Appleton is grateful for the hospitality shown by F. Mirabel and
V. Charmandaris (Paris), and C. Horellou (Onsala) in the summer of
1997, when the data reduction for the ISOCAM observations was
performed, and to F. Ghigo (NRAO) for similar hospitality at NRAO
Green Bank in 1995. We are grateful to an anonymous referee for
helpful comments on the manuscript.  The authors have enjoyed
discussions with C. Struck (ISU), J.H. Black (Onsala Space
Observatory) and F. Combes (Paris Obs.). This work is supported in
part by NASA/NAG 5-3317 (ISOCAM) and NSF grant AST-9319596 (radio
observations). V. Charmandaris would like to acknowledge the financial
support from a Marie Curie fellowship grant (ERBFMBICT960967).

\appendix

\section{Upper Limits to the Molecular and Total Gas Content
of VII\,Zw\,466 and its companion VII\,Zw\,466~(Cl):G2}

In $S$2.2 we describe the OSO observations of VII\,Zw\,466 and G2.  No
significant CO emission was detected from either of the two galaxies.
Upper limits can be determined from the noise in the final spectra
($\sigma_{mb}$=2.3 mK for VII\,Zw\,466 and 10 mK for G2) and the
\ion{H}{1} linewidths ($\Delta v_{HI}=202$\,km\,s$^{-1}$ for
VII\,Zw\,466 and 84 km\,s$^{-1}$ for G2; see ACS).  The inferred
limits on the H$_2$ masses given in Table A1 were calculated using a
standard CO to H$_2$ conversion factor (N(H$_2$)/I(CO)= 2.3\,$\times$
10$^{20}$\,mol\,cm$^{-2}$\,(K\,km\,s$^{-1}$)$^{-1}$; Strong et
al. 1988). We caution that it is not known whether this factor is
appropriate for collisional systems.

The limit on the value of log(M(H$_2$)/L$_B$) ratio for VII\,Zw\,466
is $<$ -1.46 for VII\,Zw\,466 and $<$ -0.93 for G2 (L$_B$ is based on
the CCD photometry of Appleton \& Marston 1997).  Hence VII\,Zw\,466
has a slightly lower than the average value measured in ring galaxies
but not significantly ($<$log(M(H$_2$)/L$_B$)$>$ = -1.18$\pm$0.41;
Horellou et al. 1995). Given the star formation rate of 0.97
M$\odot$/yr derived from the H$\alpha$ imaging, it would seem that
VII\,Zw\,466 would take at least 1 Gyr to deplete its molecular gas
mass, and this is longer than the dynamical time for the ring to
propagate out of the disk ($\sim$ 10$^8$ yrs--see ACS).

The results also allow us to limit the total gas to stellar
luminosity. In ACS, it was suggested that the low neutral hydrogen to
optical luminosity for G2 may be evidence for significant gas
stripping as a result of its passage through the disk of VII\,Zw\,466
in the past. The molecular-line observations provide
constraints on any molecular gas which may be present in both
galaxies. Based on the \ion{H}{1} observations of ACS, we can now
place limits of the total gas mass in both galaxies, and various gas
properties of the galaxies. These are presented in Table A1. The
results show that both VII\,Zw\,466 and G2 have total gas to light
ratios which are similar. The molecular upper limit on G2 is 
not sufficiently stringent to support the idea that G2 is gas poor
(the current limit on molecular hydrogen is a factor of two
higher than the detected \ion{H}{1} mass).

\placetable{tablea1}

%
%
 
\clearpage 
\begin{deluxetable}{lcccccc} 
\scriptsize 
\tablecaption{Mid-IR Flux Densities\label{table1}} 

\tablewidth{0pc} 
\tablehead{ 
\colhead{Name} & 
\colhead{RA(1950)\tablenotemark{a}}   & 
\colhead{DEC(1950)\tablenotemark{a}}   & 
\colhead{F$_\nu$(4.5$\mu$m) mJy} &  
\colhead{F$_\nu$(9.62$\mu$m) mJy} &  
\colhead{F$_\nu$(11.4$\mu$m) mJy} & 
\colhead{F$_\nu$(15.0$\mu$m) mJy}
}
\startdata 
VII\,Zw\,466 (Total) & 12 29 51.5 & 66 40 46 & $<$0.8 & 5.70 (1.00) & 5.52 (2.50) &  5.76 (2.50)\nl 
VII\,Zw\,466 (N Knot) &  &  & -- & 1.71 (0.25) & 1.56 (0.83)& 1.71 (0.63)\nl 
VII\,Zw\,466 (E Knot) &  &  & -- & 0.60 (0.10)& 0.31 (0.20) & 0.55 (0.25)\nl 
VII\,Zw\,466 (W Knot) &  & & --  & 1.21 (0.18) & 1.65 (0.40) & 1.24 (0.43)\nl 
G1 & 12 30 01\tablenotemark{b} & 66 40 30\tablenotemark{b} & 1.82 (0.8)& 0.1 (0.05) & $<$0.4 & $<$0.3 \nl 
G2 & 12 29 59.0 & 66 39 55 & 1.0 & 7.63 (0.56) & 8.91 (1.2) & [3.34 (1.40)]\tablenotemark{c}\nl 
B1 & 12 29 58.0 &66 40 50  &1.0  & 4.67 (0.42) & 4.08 (0.87) & 3.83 (1.00)\nl 
\enddata 
\tablenotetext{a}{from 3.8cm VLA radio positions} 
\tablenotetext{b}{from optical position (Jeske 1986)} 
\tablenotetext{c}{uncertain flux-see text} 

\end{deluxetable} 
 
\clearpage 
\begin{deluxetable}{lccccc} 
\scriptsize 
\tablecaption{Radio Fluxes, Spectral Indices and Upper Limits\label{table2}} 
\tablewidth{0pc} 
\tablehead{ 
\colhead{Name} &  
\colhead{~~~$S_{20\,cm}$ mJy} & 
\colhead{~~~$S_{6\,cm}$ mJy} & 
\colhead{~~~$S_{3.8\,cm}$ mJy} & 
\colhead{~~~$\alpha$($_{20-6\,cm}$)\tablenotemark{a}}  & 
\colhead{~~~$\alpha$($_{6-3.6\,cm}$)\tablenotemark{a}}   
}  
\startdata 
VII\,Zw\,466 & 1.81 & 0.97 & 0.69 & -0.51 & -0.63\nl 
G1\tablenotemark{b} & $<$ 0.14 & $<$ 0.02 & $<$ 0.02 & --- & ---\nl 
G2	     & 1.52  & 0.60 & 0.52 & -0.75 & -0.28\nl 
B1	     & 0.76  & 0.26  & 0.14 & -0.87 & -1.18 \nl 
\enddata 
\tablenotetext{a}{spectral index $\alpha$ of the form F$_1$/F$_2$=$\nu$$^{\alpha}$} 
\tablenotetext{b}{1$\sigma$ upper limits in units of mJy/beam area} 

\end{deluxetable} 
 
\clearpage 
\begin{deluxetable}{lccc} 
\scriptsize 
\tablecaption{Expected and Observed Mid-IR Flux Ratios\label{table3}} 
\tablewidth{0pc} 
\tablehead{ 
\colhead{Region} &  
\colhead{~~~LW8/L7} & 
\colhead{~~~LW9/LW7} & 
\colhead{~~~Comment}  
}  
\startdata 
Antenna Knot A   & 1.74 & 4.91 & Starburst region with strong thermal continuum\nl 
Antenna Knot B   & 0.91 & 0.99 & Starformation region in presence of [NeIII]\nl 
Antenna Knot Z   & 1.08 & 0.70 & Starformation region with little [NeIII]\nl 
Arp 220          & 0.32 & 6.22 & Powerful starburst with strong Silicate absorption\nl 
VII\,Zw\,466         & 0.97 & 1.01 & Ring galaxy as a whole\nl 
VII\,Zw\,466 W Knot  & 1.36 & 1.03 & Western knot in ring\nl 
Galaxy G2        & 1.17 & [0.43] & Uncertain flux in LW9 due to stripe\nl 
Galaxy B1        & 0.87 & 0.82  & Higher redshift galaxy\nl 
\enddata

\end{deluxetable} 

\clearpage
\begin{deluxetable}{lccc}
\scriptsize
\tablecaption{Optical and Mid-IR Properties of the Ring Knots\label{table4}}         

\tablewidth{0pc}
\tablehead{
\colhead{Knot Number\tablenotemark{a}} &
\colhead{~~~$S_{LW7}$ Flux mJy} &
\colhead{~~~$S_{H{\alpha}}$ Flux (x 10$^{-14}$) ergs s$^{-1}$cm$^{-2}$} &
\colhead{~~~Notes} 
}
\startdata
Knot 1+2 & 1.71 & 5.2 & northern knots\nl
Knot 3   & 0.60 & 1.5 & eastern knot\nl
Knot 4   & $<$ 0.05 & 2.3 & \nl
Knot 5+6+7 & 1.21 & 6.8 & western complex \nl
Knot 8 & $<$ 0.05 & 1.66 & north-western knot \nl
\enddata
\tablenotetext{a}{Knots defined in Figure 2a}

\end{deluxetable}  

\clearpage
\begin{deluxetable}{lccccccc}
\tablenum{A1}
\scriptsize
\tablecaption{Total Gas Properties of VII\,Zw\,466 and VII\,Zw\,466:G2\label{tablea1}}   
\tablewidth{0pc}
\tablehead{
\colhead{Name} 				&
\colhead{M(HI)} 			& 
\colhead{M(H$_2$)} 			& 
\colhead{M(HI)/M(H$_2$)} 		&
\colhead{M(HI+H$_2$)} 			& 
\colhead{L$_B$}          		& 
\colhead{log(M(H$_2$)/L$_B$)}		&
\colhead{log(M(HI+H$_2$)/L$_B$)}	\\
\colhead{}				&
\colhead{10$^9$ M$_{\odot}$}		&
\colhead{10$^9$ M$_{\odot}$}		&
\colhead{}				&
\colhead{10$^9 $M$_{\odot}$}		&
\colhead{10$^{10}$ L$_\odot$}		&
\colhead{M$_{\odot}$/L$_{\odot}$}	&
\colhead{M$_{\odot}$/L$_{\odot}$}}
\startdata
VII\,Zw\,466 & 4.1 & $<$1.2  & $<$0.29 & 4.1--$<$5.3 & 3.44 &  $<$-1.46 & $<$-0.81 \nl
VII\,Zw\,466:G2 & 1.2 & $<$2.1 & $<$ 1.75 & 1.2--$<$3.3 & 1.82 &$<$-0.93
& $<$ -0.74 \nl
\enddata

\end{deluxetable}

%
%

\newpage 

%
%
 
\clearpage 

\clearpage

\newpage

\end{document}